\documentclass{svjour2}
\usepackage{graphicx}
\usepackage{rotating}
\usepackage{amssymb}
\usepackage{mathptmx}
\usepackage{authblk}
\usepackage[numbers]{natbib}
\usepackage[]{units}
\makeatletter
\journalname{Journal of Low Temperature Physics}

\bibpunct{}{}{,}{s}{}{,}

\begin{document}

\newcommand{\hdblarrow}{H\makebox[0.9ex][l]{$\downdownarrows$}-}
\title{Latest Progress on the QUBIC Instrument}

\author{A. Ghribi$^1$ \and J. Aumont$^4$ \and E. S. Battistelli$^6$ \and A. Bau$^7$ \and B. B\'elier$^{14}$  \and L. Berg\'e$^3$ \and J-Ph. Bernard$^2$ \and M. Bersanelli$^8$ \and M-A. Bigot-Sazy$^1$ \and G. Bordier$^1$ \and E. T. Bunn$^{12}$ \and F. Cavaliere$^8$ \and P. Chanial$^1$ \and A. Coppolecchia$^6$ \and T. Decourcelle$^1$ \and P. De Bernardis$^6$ \and M. De Petris$^6$ \and A-A. Drilien$^3$ \and L. Dumoulin$^3$ \and M. C. Falvella$^{13}$ \and A. Gault$^{11}$ \and M. Gervasi$^7$ \and M. Giard$^2$ \and M. Gradziel$^5$ \and L. Grandsire$^1$ \and D. Gayer$^5$ \and J-Ch. Hamilton$^1$ \and V. Haynes$^9$ \and Y. Giraud-H\'eraud$^1$ \and N. Holtzer$^3$ \and J. Kaplan$^1$ \and A. Korotkov$^{10}$ \and J. Lande$^2$ \and A. Lowitz$^{11}$ \and B. Maffei$^9$ \and S. Marnieros$^3$ \and J. Martino$^4$ \and S. Masi$^6$ \and A. Mennella$^8$ \and L. Montier$^2$ \and A. Murphy$^5$ \and M. W. Ng$^9$ \and E. Olivieri$^3$ \and F. Pajot$^4$ \and A. Passerini$^7$ \and F. Piacentini$^6$ \and M. Piat$^1$ \and L. Piccirillo$^9$ \and G. Pisano$^9$ \and D. Pr\^ele$^1$ \and D. Rambaud$^2$ \and O. Rigaut$^3$ \and C. Rosset$^1$ \and M. Salatino$^6$ \and A. Schillaci$^6$ \and S. Scully$^5$ \and C. O'Sullivan$^5$ \and A. Tartari$^1$ \and P. Timbie$^{11}$ \and G. Tucker$^{10}$ \and L. Vibert$^4$ \and F. Voisin$^1$ \and B. Watson$^9$ \and M. Zannoni$^7$}

\institute{$^1$AstroParticule et Cosmologie, Univ. Paris 7, CNRS \and $^2$Institut de Recherche en Astrophysique et Plan\'etologie \and $^3$Centre de Spectrom\'etrie Nucl\'eaire et de Spectrom\'etrie de Masse \and $^4$Institut d'Astrophysique Spatiale\and $^5$NUI Maynooth \and $^6$Universit\`a di Roma Ð La Sapienza \and $^7$Universit\`a di Milano Ð Bicocca \and $^8$Universit\`a degli studi Milano \and $^9$University of Manchester \and $^{10}$Brown University \and $^{11}$UniversityÊof Wisconsin \and $^{12}$University of Richmond \and $^{13}$Italian Space Agency \and $^{14}$Institut  d'Electronique Fondamentale \\ \\
\email ghribi@in2p3.fr}

\date{\today}

\maketitle

\begin{abstract}

QUBIC is a unique instrument that crosses the barriers between classical imaging architectures and interferometry taking advantage from both high sensitivity and systematics mitigation. The scientific target is to detect primordial gravitational waves created by inflation by the polarization they imprint on the Cosmic Microwave Background -- the holy grail of modern cosmology. In this paper, we show the latest advances in the development of the architecture and the sub-systems of the first module of this instrument to be deployed at Dome Charlie Concordia base - Antarctica in 2015.


\end{abstract}

\section{Introduction}

The Cosmic Microwave Background (CMB) has been heavily observed and studied in the last decades. From this keen interest emerged a number of experiments, backed by important R\&D efforts, that put tight constraints in favour of inflationary cosmology models. One last prediction, the presence of primordial gravitational waves resulting in tensor mode perturbations in the metric, is still undetected. These space-time fluctuations would have left a faint but indelible trace on the polarization of the CMB photons\cite{zaldarriaga2004}. Today, thanks to the development of large arrays of photon noise limited detectors, this detection seems to be within reach. One last barrier could prevent us from detecting it: uncontrolled systematic effects. In the landscape of imaging instruments, usually suffering from uncontrolled time varying systematics (e.g. atmospheric contamination), one outsider architecture was proposed in the early 2000: bolometric interferometry\cite{ali, piccirillo}. This new architecture is now making its way through theoretical and technological developments: bolometric interferometry\cite{dibo, charlassier}. It uses background limited incoherent detectors but takes advantage of baseline redundancy to make interferometry a viable option without significant sensitivity loss\cite{qubic2010} thanks to a special self calibration technique\cite{marieanne}. In this type of architecture, equivalent baselines are coherently summed on every detector avoiding the need for expensive, complex and noisy correlators, mixers and amplifiers. QUBIC (Q and U Bolometric Interferometer for Cosmology) is today the only instrument based on bolometric interferometry.

In this paper, we report the latest advances on the development of the 1st module of the QUBIC instrument. After introducing  the QUBIC instrument, we will review progress on each of its subsystems.

\section{The QUBIC Instrument}

QUBIC will ultimately be composed of 6 single-frequency modules, operating at \unit[97]{GHz}, \unit[150]{GHz} and \unit[220]{GHz} with 25\% bandwidth. Each module will respectively comprise 144, 400 and 625 horns. The first module will be centered around 150GHz as the base CMB channel, taking advantage of recently released data (Planck HFI) for foreground removal and a unique self calibration strategy to mitigate instrumental systematic effects\cite{marieanne}. 
For that purpose, it will use electromagnetic switches sandwiched between the primary and the secondary horn arrays. The primary horns look at the sky through optical filters and a cold rotating half wave plate. The secondary horn array re-emits the signal to illuminate a beam combiner. The latter realizes the necessary path-dependant phase shift to allow coherent summation of equivalent baselines on the focal plane after polarization separation. The focal planes will be made of two arrays of 1024 TES bolometers each cooled to a temperature of \unit[0.3]{K}. 
Currently the subsytems are being fabricated and tested. First light at the observation site is scheduled for 2015. The expected sensitivity to the tensor to scalar ratio after one year of observation is 0.05 with one module and 0.008 with 6 modules\cite{qubic2010}. The observation site will be at Dome Charlie Concordia Station in Antarctica, a site that has proven to be one of the best on earth for CMB observations\cite{batistelli}. Previous campaigns at Dome C have already allowed the characterization of the atmosphere as well as the improvement of the logistics\cite{polenta,batistelli}.

\section{First Module Instrument Subsystems}

\subsection{Cryogenics}
\begin{figure}

\includegraphics[width=11.5cm]{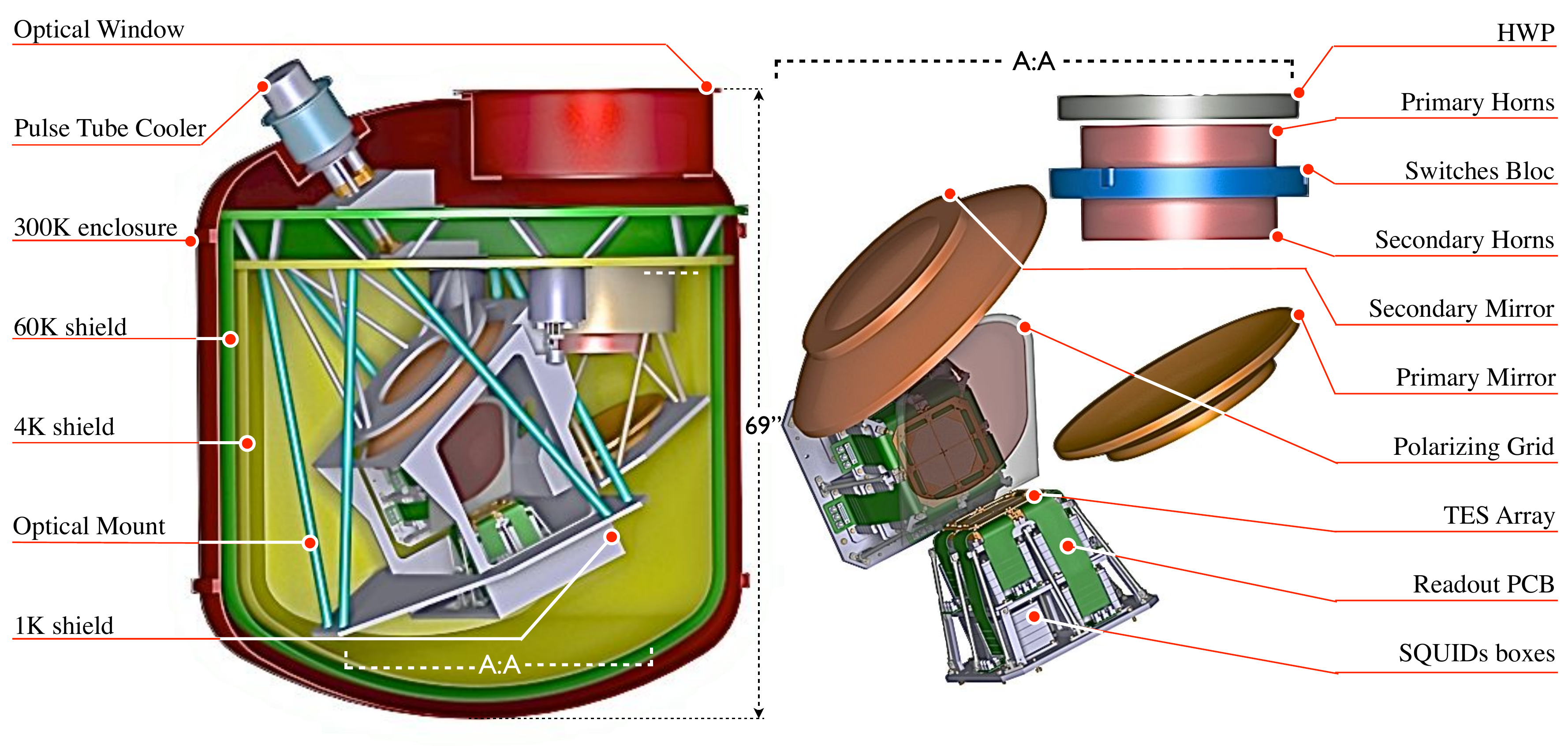}
\centering{}
\caption{Left: cryostat sketch design. Right: optical system, focal plane and readout blocks. (colour figure online)}\label{fig:cryostat}
\end{figure}

The large size of the horn array, the optical combiner and the focal plane drives the overall dimensions of the cryostat (see figure \ref{fig:cryostat}). The cold volume is of the order of 1.3 m$^3$, and a large (60 cm in diameter) window is required by the large optical throughput of the system. All-aluminum construction \cite{masietal} is necessary to limit the weight of the system and the requirements on the mount. For operation in the harsh Dome-C environment, where supply of cryogenic liquids is difficult, we opted for a dry cryostat cooled by a Sumitomo pulse tube refrigerator, adapted for operation at low temperatures in Dome-C by means of suitable heaters\cite{polenta}. The detector arrays are cooled to \unit[0.3]{K} by a $^3$He sorption cooler, while the cold optics are cooled below \unit[4]{K} by a $^4$He sorption cooler.

\subsection{Optics}
\subsubsection{Filters and Half Wave Plate}

Metal mesh interference filters reduce the radiative load on the cryogenics and detectors and, in combination with a section of waveguide between each back-to-back primary/secondary horn, define the spectral band. A Half Wave Plate (HWP) will be used at \unit[4]{K} in order to mitigate systematic effects and decrease the 1/f noise. Measurements of the HWP are in good agreement with the modeled performance, with a differential phase error of a few degrees and cross-polarization at the \unit[- 35/-40]{dB} level with a bandwidth of ~ 30\%\cite{pisano}. The rotating mechanism is a scaled-up version of the one developed for the PILOT experiment\cite{salatino}. It will allow a stepped movement with a position control better than \unit[0.1]{deg}.


\subsubsection{Horn Array} 
The first module will have a close-packed array of 400 primary horn antennas backed by a similar array of secondary antennas, sandwiching an array of aperture shut-off switches.. These single-mode corrugated horn antennas operate around \unit[150]{GHz} with a beam of \unit[14]{deg} FWHM and have low cross-polar response. They will be made of a stack of aluminum platelets mounted with an accuracy better than \unit[30]{$\mu$m}. At \unit[90]{GHz} this fabrication technique has proven to be very efficient and has produced horns with good agreement with simulations.\cite{petris}.

\subsubsection{Switches}
During calibration (but not sky observations), an array of 400 switches will be used as shutters, operated independently for each channel. They will be operated by magnetically shielded electromagnets, controlled and de-multiplexed by a cold ASIC\footnote{Application Specific Integrated Circuit}. Measurement of the inductance of each coil will allow cross-verification of the state of the switches. These devices have been successfully tested at \unit[77]{K} and show a power dissipation of \unit[150]{mW}. Only two switches are activated at a time during the self-calibration procedure. The switches are not expected to affect the magnetically shielded SQUIDs\footnote{Superconducting Quantum Interference Devices} (see section 3.3.2).

\subsubsection{Beam Combiner}
Beam combination in QUBIC is performed by means of an optical system that propagates the fields radiated by the back-to-back horn array to a detector plane where Fizeau interference fringes can be observed.  The beam combiner\cite{combiner} is an off-axis Gregorian dual reflector, designed using the Mizuguchi-Dragone condition to have low spherical aberration, astigmatism and geometrical cross polarization across as large a field-of-view as possible. A polarizing grid separates the signal into two orthogonal polarizations that are imaged onto separate detector grids and the entire combiner fits inside a $\sim \unit[1.3]{m^3}$ cold volume (see figure \ref{fig:cryostat}). A field-of-view of 14deg together with a focal length and  entrance aperture diameter of \unit{300}[mm] makes the design extremely challenging but simulations have shown that the effect of aberrations in our design is to reduce the overall sensitivity, compared to an ideal combiner, by approximately 10\%.

\subsection{Detection Chain}

\subsubsection{Detectors}
\begin{figure}
\includegraphics[width=6.6cm]{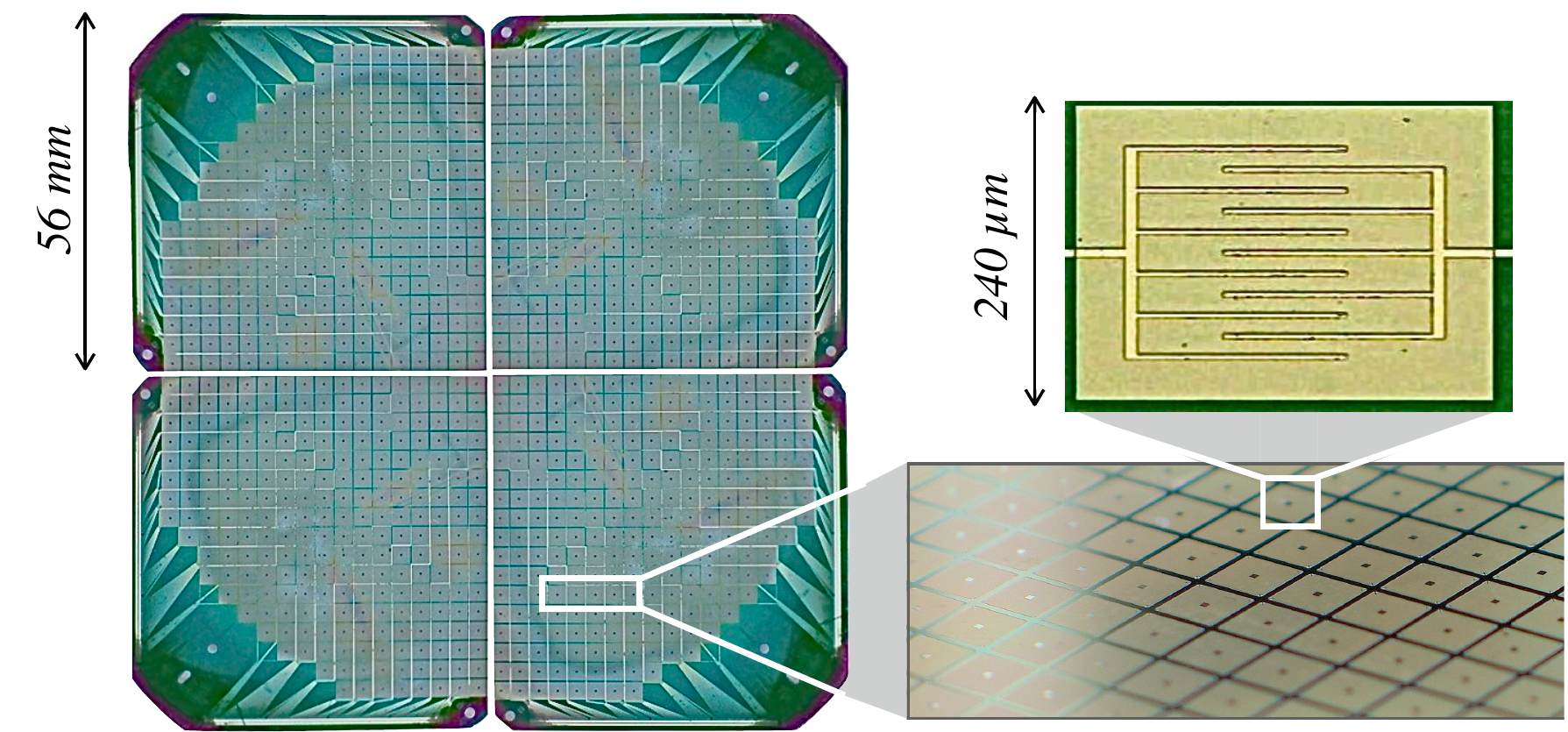}
\centering{}
\caption{Picture of preliminary realizations of detector sub-arrays with a zoom on the thermistor.}\label{fig:array}
\end{figure}

The focal plane will be covered with two filled arrays in order to Nyquist sample the image of the interference fringes. Each array will detect a linear polarization either transmitted or reflected by a polarized grid. The 1024 detectors of each array  (figure \ref{fig:array}) are based on \unit[0.3]{K} cooled Transition Edge Sensors with a target noise equivalent power (NEP) of \unit[$4~10^{-17}]{\unitfrac{W}{\sqrt{Hz}}}$ and a time constant shorter than \unit[20]{ms}. A single bolometer is made of an island of suspended $TiV$ absorber on a low stress $SiN_x$ membrane heating an $Nb_xSi_{1-x}$ thermistor. Four $SiN_x$ legs provide the thermal link to the bulk silicon. This design has the advantage of independently tuning the critical temperature by varying the concentration of $Si$ and $Nb$ in the $Nb_xSi_{1-x}$ alloy and the normal resistance by tuning the geometry of $Nb$ interdigitated electrodes\cite{pajot} (see figure \ref{fig:array}). A target Tc of \unit[0.4]{K} and normal resistance of \unit[200]{m$\Omega$} has been set to reach the required NEP and time constant. These parameters result from a tradeoff between the required sensitivity, the saturation power to observe the artificial calibration source and the capabilities of the cryogenic system that will be used for the first module.
A first array is being fabricated and is scheduled for completion in September 2013. In the meanwhile, 23 pixel detectors arrays have been characterized and show an NEP of \unit[$4~10^{-17}]{\unitfrac{W}{\sqrt{Hz}}}$ for a Tc of \unit[0.5]{K} \cite{joseph}.

\subsubsection{Readout Electronics}
The readout electronics use SQUIDs as a first time-domain multiplexing stage followed by a cryogenic full custom BICMOS SiGe ASIC. The SQUIDs are integrated with second order gradiometers and each one of them is enclosed in a separate tin-coated ferromagnetic box that provides magnetic shielding.
The ASIC controls the capacitive SQUID addressing for row readout and second stage amplification. The advantages of such a system are a simplification of the architecture, miniaturization of the readout electronics, low power consumption and very low noise properties. The ASIC uses a single voltage supply of \unit[3.3]{V} and consumes only \unit[15]{mW} per ASIC. This technique has been demonstrated\cite{damien} with a 24 pixel readout in a 24:1 multiplexing scheme (done in two steps: 8:1 by the SQUID stage and 3:1 with the ASIC). The current design is based on a new ASIC that is able to read out 128 detectors with a 128:1 multiplexing factor (32:1 SQUIDs and 4:1 ASIC). The full detector array will therefore be read out with 2048 SQUIDS and 16 SiGe ASICs.

Data acquisition and control software and warm electronics design and fabrication are ongoing at IRAP Toulouse. These components are designed to facilitate operation during the Antarctic winter.

\section{Conclusion}
Significant progress has been made in the design, fabrication and characterization of the subsystems for the first module of QUBIC. Its deployment at Dome C will allow us to achieve a tensor to scalar ratio of 0.05 within one year of observation. It will also allow the demonstration of the importance of the self calibration strategy to reach unprecedented levels of systematics control in CMB instruments. The 5 modules that will follow will be a real breakthrough with a target tensor to scalar ratio lower than 0.01. For future modules several improvements will be considered: A dilution fridge will allow to achieve lower temperature and hence improve TESs performance. Planar orthogonal mode transducers will give better optical performance than a polarizing grid, reducing the number of focal planes from two to one. Superconducting switches will allow very fast, low power consumption and compact baseline shut-off for the self calibration. 
It is important to note that the first QUBIC module is not only a scientific demonstration on CMB photons of the capabilities of bolometric interferometry but also a technology demonstration of new techniques for a future cosmology space mission.

\begin{acknowledgements}
Agence Nationale de la Recherche (ANR), Centre National dEtudes Spatiales (CNES), Centre National de la Recherche Scientifique (CNRS), Science and Technology Facilities Council (STFC), the National Science Foundation (NSF), the National Aeronautics and Space Administration (NASA), Instititut Paul Emile Victor (IPEV), Programma Nazionale Ricerche in Antartide (PNRA), Science Foundation Ireland (SFI). and Labex UnivEarthS.
\end{acknowledgements}



\begin{thebibliography}{99}


\bibitem{zaldarriaga2004}
Zaldarriaga M., Carnegie Obs. astro-ph.SR, 2, 309, 2004.

\bibitem{qubic2010}
Batistelli E. et al., Astropart. Phys., 34, Issue 9, p. 705-716. 2011.

\bibitem{masietal}
Masi S., et al., Cryogenics, 39, 217-224, 1999.

\bibitem{polenta}
Polenta G., et al., New Astronomy Reviews, 51, 256-259, 2007.

\bibitem{batistelli}
Battistelli E. S. et al., Mon. Not. R. Astron. Soc. 423, 1293Ð1299, 2012.


\bibitem{ali}
Ali S. et al., AIP conf. proc., 616, pp.126-128, 2001.

\bibitem{piccirillo}
Piccirilo L., Mem. Soc. Astr. It. S., 2,p200, 2003.

\bibitem{dibo}
Ghribi A. et al., J. Infrared Millim. Te., 31, 1, pp.88-99. 2010.

\bibitem{charlassier}
Charlassier et al., A\&A, vol. 497, 3, 2009, pp.963-971, 2009.

\bibitem{marieanne}
Bigot-Sazy M-A., A\&A, 550, id.A59, 11 pp., 2013.

\bibitem{salatino}
M. Salatino, P. de Bernardis, S. Masi, A\&A, 528, A138, 2011.

\bibitem{pisano}
Pisano G. et al. PIER M, 25, 101,2012.

\bibitem{petris}
Torto F. Del et al. JINST,6, P06009, 2011.

\bibitem{pajot}
Pajot F. et al., J. Low Temp. Phys., 151, 513, 2008.

\bibitem{joseph}
Martino J., PhD thesis, 2012.

\bibitem{damien}
Pr\^ele D. et al., European Astronomical Society, 37, 2009.

\bibitem{combiner}
Gayer D. et al., Proc. SPIE, 8452, 8 pp., 2012.

\end{thebibliography}
\end{document}